\documentclass[runningheads]{cl2emult}

\usepackage{makeidx}  
\usepackage{graphicx} 
\usepackage{subeqnar} 
\usepackage{multicol} 
\usepackage{cropmark} 
\usepackage{eso}      
\makeindex            

\providecommand*{\dfrac}[2]{\displaystyle\frac{#1}{#2}}
\begin{document}
\title*{The 2dF QSO redshift survey}
%
%
%
%
\titlerunning{The 2dF QSO redshift survey}
%
\author{B.J.Boyle\inst{1}
\and S.M.Croom\inst{1}
\and R.J.Smith\inst{2}
\and T.Shanks\inst{3}
\and P.J.Outram\inst{3}
\and F.Hoyle\inst{3}
\and L.Miller\inst{4}
\and N.S.Loaring\inst{4}
}
\authorrunning{B.J.Boyle et al.}
%
%
\institute{Anglo-Australian Observatory, PO Box 296, Epping, 
NSW 1710, Australia
\and 
Liverpool John Moores University, Twelve Quays House, Egerton Wharf,
Birkenhead, CH41 1LD, UK
\and
Department of Physics, University of Durham, South Road, Durham, 
DH1 3LE, UK
\and
Department of Physics, University of Oxford, 1 Keble Road,
Oxford, OX1 3RH, UK
}

\maketitle              

\begin{abstract}
We present some initial results from the 2dF QSO redshift survey.
The aim of the survey is to produce an optically-selected catalogue of 
25000 QSOs over the redshift range $0<z<3$ using the 2-degree field
at the Anglo-Australian Telescope. 
\end{abstract}

\section{Introduction}
In the past, statistical studies of QSO clustering have been limited by
the small numbers of QSOs available in homogeneously-selected
catalogues\cite{cs96}.  The aim of the 2dF QSO redshift survey (2QZ)
is to provide a optically-selected catalogue of 25000 QSOs,
approximately 50 times larger than the previous largest QSO survey to
a similar depth ($B<21\,$mag).  This survey will be used to provide
measures of:
\begin{itemize}
\item the evolution of QSO clustering over the redshift range $0<z<3$
\item large-scale structure over scales 1--1000h$^{-1}$Mpc 
\item the cosmological parameters ($\Omega_\Lambda$, $\Omega_{\rm m}$) using
geometrical tests
\item the QSO luminosity function and its evolution with redshift
\item the spectral properties of a  large sample of QSOs
\end{itemize}

\section{Catalogue selection and observations}

\begin{figure}
\centering
\includegraphics[width=0.6\textwidth,angle=90]{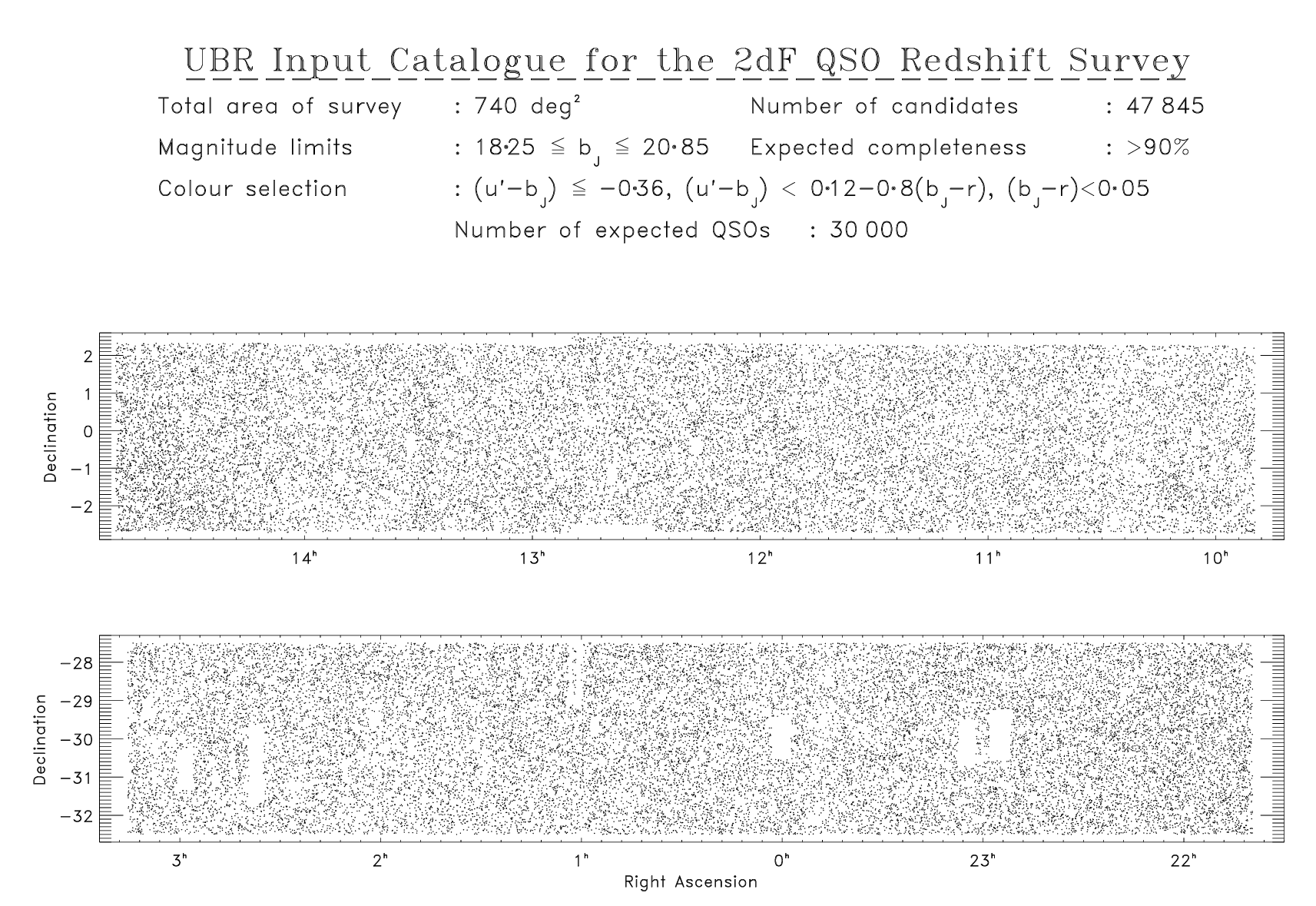}
\caption{Final QSO candidate catalogue based on $U-B_J$/$B_J-R$ selection.
Reproduced from Smith et al. (2001a).}
\label{fig:cat}
\end{figure}

The 2QZ is based on $U-B_J$/$B_J-R$ colour selection of blue candidate
objects from the Automated Plate Measuring (APM) machine scans of UK
Schmidt $U$, $J$ and $R$ plates.  Full details of the catalogue
selection process are given by Smith et al.\ (2001a)\cite{s01a}. 
The survey covers 740 deg$^2$, comprising
two $75^{\circ}\times5^{\circ}$ strips (excluding the regions around bright
stars, etc) in the South Galactic Pole region at declination $-30^{\circ}$
and in the North Galactic Cap region at declination $0^{\circ}$ (see 
Fig.~\ref{fig:cat})
  
\begin{figure}
\centering
\includegraphics[width=0.6\textwidth]{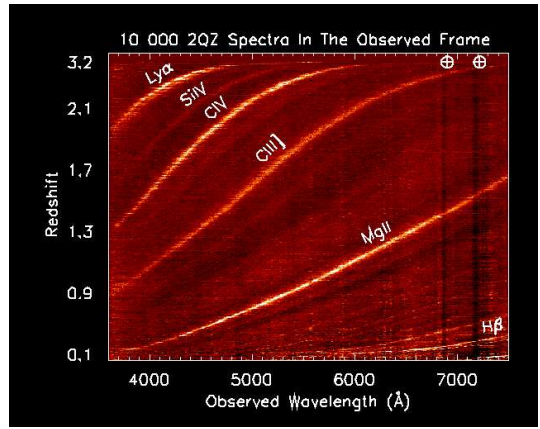}
\caption{Observed frame spectra of the first 10000 QSOs in the 2QZ.  
Spectra are stacked in order of increasing redshift. 
Reproduced from Smith et al.\ (2001b).}
\label{fig:spec}
\end{figure}

In total 47845 candidates were selected for observations with the 2dF
instrument at the Anglo-Australian Telescope.  The survey observations
are combined with those of the 2dF galaxy redshift survey 
(2dFGRS\cite{c99}).  QSO candidates are observed at 7\AA\
resolution over the wavelength range 3800--7250\AA.  We obtain
reliable identification for $\sim 85\%$ of the candidates to our survey
magnitude limit $B_J<20.85\,$mag.  Spectra for the first 10000 QSOs
observed are shown in Fig.~ref{fig:spec}.
At the time of writing (November 2000), we have observed over 32000
QSO candidates, resulting in the identification of more than 15000 QSOs.

Galactic stars represent the largest single contaminant in the
survey (25\%), including almost 1500 white dwarfs. 
Narrow emission line galaxies ($0<z<0.6$) comprise a
further 10\% of the sample.  The spectroscopic catalogue will be
released to the community in two stages.  First, an initial release of
20000 objects (including 10000 QSOs) will be made by June 30, 2001.
The release will include positions, redshift, spectra and a detailed
coverage map of the survey.  The final catalogue release will be made
one year after the survey is completed in February 2002.  The data
will be available from the 2QZ catalogue home page {\tt
http://www.2dfquasar.org}.  The input catalogue is already available
from this site.

\section{Preliminary Results}
\subsection{QSO Spectra}

\begin{figure}
\centering
\includegraphics[width=0.6\textwidth,angle=270]{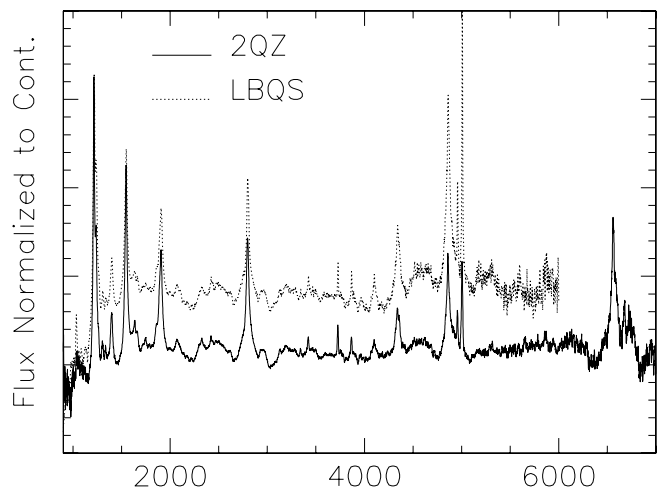}
\caption{Composite spectra for the 2QZ (lower) and LBQS (upper) samples.}
\label{fig:comp}
\end{figure}

The composite spectrum based on the co-addition of the first 10000 QSO
spectra is shown in Fig.~\ref{fig:comp}.  Full details of this work will be presented
in Smith et al. (2001, in preparation).  

The composite based on the 1000 QSOs
identified the Large Bright QSO survey (LBQS)\cite{f91} is
shown for comparison.  The LBQS composite is based on QSOs that are,
on average, 10 times more luminous than those identified in the 2QZ.
However, the two composites are almost identical.  In particular, 
the equivalent widths of the broad
emission lines (including CIV) are the same between the two surveys,
providing further evidence that the `Baldwin' effect is weak or
non-existent in optically-selected QSO catalogues\cite{z92}.  
The only difference in the emission line properties occurs
for the narrow lines; the LBQS composite exhibiting much stronger
narrow [OII] and [OIII] emission compared to the 2QZ composite.  
Given the correlation between host mass and QSO luminosity\cite{m99}, 
this suggests that the size/luminosity 
of the narrow line region is simply proportional to the mass of the host 
galaxy.

\subsection{QSO Luminosity Function}

\begin{figure}
\centering
\includegraphics[width=0.75\textwidth]{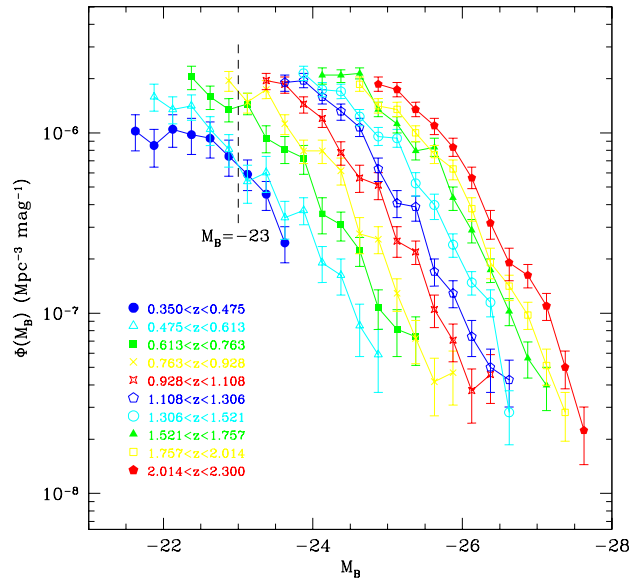}
\caption{Binned estimate of the $0.35<z<2.3$ QSO optical luminosity 
function derived from the 2QZ and LBQS samples. Reproduced from Boyle
et al.\ (2000).}
\label{fig:olf}
\end{figure}

The QSO luminosity function (LF) based on the first 5000 QSOs
identified in the 2QZ and the 1000 QSOs in the LBQS is shown in 
Fig.~\ref{fig:olf}  Full details of the analysis are given by Boyle et
al. (2000)\cite{b00}.  

For an $\Omega_{\rm m}=1$ Universe, the QSO LF
is given by a two power law model:
\[
\Phi(L,z)=\dfrac{\Phi^*}{\left[\left(\dfrac{L}{L^*(z)}\right)^{\alpha}+
\left(\dfrac{L}{L^*(z)}\right)^{\beta}\right]},
\]
\noindent
where there parameter values lie in the range 
$3.45<\alpha<3.75$, $1.55<\beta<1.96$, $-23<M_B^*<-22$ and 
$\Phi^*=10^{-6}$Mpc$^{-3}$mag$^{-1}$.  The evolution of the 
LF over the range $0.35<z<2.3$ can be fitted by a pure luminosity 
evolution model of the form:
\[
L^*(z)\propto L^*(0)10^{1.40z-0.27z^2}.
\]
This form of evolution is consistent with the $L(z)\propto (1+z)^3$
form previously derived for the redshift range $0.3<z<2$.  However the
exponential form of the relation provides for a slow down in the 
rate of PLE at $z>2$; peaking at $z=2.5$, beyond the 80\% completeness
limit for the survey.

\subsection{QSO correlation function}

\begin{figure}
\centering
\includegraphics[width=0.60\textwidth]{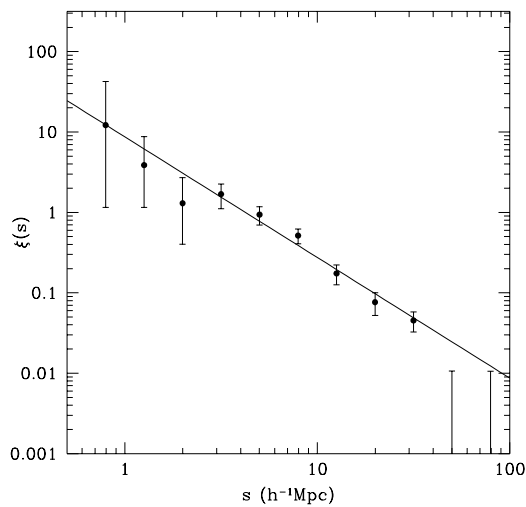}
\caption{Estimate of the mean QSO correlation function
function over the redshift range $0.35<z<2.9$ ($\Omega_{\rm m}=1)$.  
Reproduced from Croom et al.\ (2001).}
\label{fig:xi}
\end{figure}

The QSO correlation function based on the first 8000 QSOs
in the 2QZ is shown in Fig.~\ref{fig:xi}.  
Full details of the analysis are presented 
in Croom et al. (2001)\cite{c01}.  

At $r<20$h$^{-1}$ Mpc the correlation 
function, $\xi(r)$, determined over the full 2QZ redshift
range $0.35<z<2.9$ is a good fit to a power law of the form;
\[
\xi(r)=\left(\dfrac{r}{r_0}\right)^{-\gamma},
\]
\noindent
where the $\gamma=1.5$ and $r_0=4.5\pm1.0$h$^{-1}$Mpc ($\Omega_{\rm
m}=1$) at a mean redshift $z=1.4$.  This is comparable to the
present-day clustering of optically-selected galaxies.  The
clustering scale length $r_O$ shows little or no dependence on 
redshift when the sample is split into smaller redshift intervals.

For a flat Universe with $\Omega_\Lambda=0.7$, the mean
scale length over the redshift range $0.35<z<2.9$ increases to
$r_0=5.9\pm1.0$h$^{-1}$Mpc, and exhibits a strong evolution with
redshift, with $r_0$ increasing towards high redshifts.

Biasing models for QSO formation based on the
Press-Schechter formalism\cite{m97} are consistent with
the observed evolution in $r_0$ for halo masses of
$10^{12}$M$_{\odot}$ ($\Omega_{\rm m}=1$) and $10^{13}$M$_{\odot}$
($\Omega_{\rm m}=0.3$, $\Omega_{\Lambda}=0.7$).  Comparison with CDM
models yields values for the shape parameter $\Gamma\sim0.2$
($\Omega_{\rm m}=1$), $\Gamma\sim0.1$ ($\Omega_{\rm m}=0.3$,
$\Omega_{\Lambda}=0.7$).

\subsection{QSO power spectrum}

We have also determined the QSO power spectrum from the first $\sim
6000$ QSOs in the survey (Hoyle et al. 2001, in preparation).  

\begin{figure}
\centering
\includegraphics[width=0.70\textwidth]{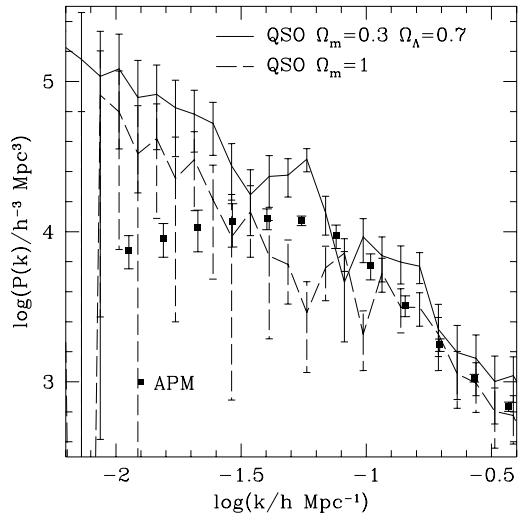}
\caption{The QSO power spectrum obtained from the first
6000 QSOs in the 2QZ, shown for both a $\Omega_{\rm m}=1$ and
$\Omega_{\rm m}=0.3$, $\Omega_{\Lambda}=0.7$ Universe.
The APM P(k) from Baugh \& Efstathiou (1993) is
also shown. Reproduced from Hoyle et al., in preparation.}
\label{fig:pk}
\end{figure}

Over scales $20-150$h$^{-1}$Mpc, the QSO redshift-space power spectrum
(see Fig.\ref{fig:pk}) shows a similar shape, $P(k)\propto k^{-1.6}$,
to that measured for galaxies and clusters at low redshift.  The
amplitude of the QSO power spectrum is similar to the present epoch
galaxy power spectrum for $\Omega_{\rm m}=0.3$, $\Omega_{\Lambda}=0.7$
and a factor of 2 lower for $\Omega_{\rm m}=1$.  At larger scales, the
QSO power spectrum displays excess power over that derived from the
APM galaxy survey\cite{be93} Fitting CDM models to the observed QSO
spectrum yields low values of $\Gamma$ consistent with those obtained from
the analysis of the correlation function.

\section{Conclusions}

The 2QZ is now over 60\% complete.  It has already begun to provide
important new information on the spectral properties, luminosity
function, evolution and clustering of QSOs.  The initial catalogue of
10000 QSOs will be publically released by June 30, 2001.  Combined
with other major QSO surveys (SDSS, FIRST), the 2QZ should result in a
much greater understanding of both the QSO phenomenon and the
evolution of large-scale structure in the early Universe.

\end{document}